\documentclass[slac_one]{revtex4}
\usepackage{graphicx}
\usepackage{fancyhdr}
\begin{document}

\title{Standard Model Higgs Searches at the LHC} %% Paper title goes here

\author{R.~Gon\c{c}alo}
\affiliation{Royal Holloway, University of London}

\author{On behalf of the ATLAS and CMS collaborations}

\begin{abstract}
  The study of the mechanism behind electroweak symmetry breaking is
  one of the main goals of the Large Hadron Collider and of its
  general-purpose experiments, ATLAS and CMS. This paper reviews some
  of the ongoing studies by these collaborations and, when possible,
  highlights the differences between equivalent channels in both
  experiments. 
\end{abstract}

%\maketitle must follow title, authors, abstract
\maketitle

\thispagestyle{fancy}

% body of paper here - Use proper section commands
% References should be done using the \cite, \ref, and \label commands
% Put \label in argument of \section for cross-referencing
%\section{\label{}}

\section{Introduction} % Section title should be in all capitals.

The Higgs mechanism~\cite{ref:higgs} is central to the Standard Model
of particle physics (SM). The existence of the Higgs field maintains
the theory weakly interacting up to high energy scales and prevents
some processes from violating unitarity. In the process of spontaneous
breaking of electroweak symmetry, the weak bosons $W^{\pm}$ and $Z^0$,
as well as leptons and quarks, acquire mass through interactions with
the Higgs field. In the SM, the simplest form of the Higgs mechanism
is assumed, which predicts the existence of a single scalar particle,
the Higgs boson and a single free parameter, its mass ($m_H)$. The
discovery of this particle would provide experimental evidence for the
Higgs mechanism. The discovery and study of the mechanism of
electroweak symmetry breaking is one of the main goals of the
ATLAS~\cite{ref:detpaper} and CMS~\cite{ref:cmstdr1} experiments,
operating at the Large Hadron Collider (LHC).  Many models of physics
beyond the SM predict a more complex Higgs sector, covered
elsewhere~\cite{ref:lowette}.

Direct searches for the Higgs boson produced in association with a
$Z^0$ were performed in the Large Electron Positron collider
(LEP). These resulted in the exclusion of the Higgs boson in the mass
range up to 114.4~GeV at 95\% confidence level~\cite{ref:lephwg03}. On
the other hand, precision fits~\cite{ref:lepewwg08} of electroweak
observables, including data from the LEP and Tevatron colliders,
provide an indirect estimate of the Higgs boson mass, assuming the SM
scenario.  The latest fit results give $m_{H} = 84^{+34}_{-26}
\,\rm{GeV}/c^{2}$, or the one-sided 95\% confidence-level limit $m_{H}
< 154 \, \rm{GeV}$. Including the LEP direct search results, this
limit increases to $185\,\rm{GeV}/c^2$. Recent combined results from
the Tevatron experiments have, for the first time, excluded the
hypothesis of a Higgs boson mass around $170 \,
\rm{GeV}$~\cite{ref:herndon} at 95\% confidence level.  Although the
expected sensitivity of Tevatron experiments is not enough to make a
5$\sigma$ discovery of the SM Higgs boson~\cite{ref:rembold}, it is
enough to exclude it out up to $m_{H} \sim 200 \, \rm{GeV}/c^2$ at
95\% confidence level, or to make a $3\sigma$ observation.

\section{Higgs boson searches in ATLAS and CMS}

The Higgs boson production at the LHC is dominated by the gluon-gluon
fusion process, described at leading order through a heavy-quark loop.
The next-to-leading order cross section for this process is $37.6
\,pb$, for $m_H=120\,\rm{GeV}/c^2$. The Higgs boson can also be
produced by Vector Boson Fusion (VBF) with a cross section of $4.25
\,pb$, or by associated production with a $W^{\pm}$, a $Z^0$, or a
$t\bar{t}$ quark pair, with $3.19 \,pb$ for the three processes and
$m_H=120\,\rm{GeV}/c^2$ (cross sections calculated at next-to-leading
order using parton density function sets CTEQ6M and
CTEQ6L1~\cite{ref:cscbook}).

The Higgs boson branching ratio is strongly dependent on its mass. At
$m_H \lesssim 135 \, \rm{GeV}/c^{2}$, the main decay mode is to a
$b\bar{b}$ pair ($BR = 81\%$), followed by the decay to a
$\tau^+\tau^-$ pair ($BR\sim 8\%$). For a small but important interval
of $m_{H}$, the Higgs boson decays to a pair of photons with a small
branching ratio. At higher masses, the decay to a pair of (possibly
off-shell) $W^{\pm}$ or $Z^0$ bosons becomes dominant.

The most abundant signal topologies, containing $b\bar{b}$ pairs are
unfortunately hard to separate from the large QCD background.  The
following gives a summary of the discovery channels being investigated
at the LHC.  In all cases, full detector simulation was used, which
included realistic descriptions of the material budget and detector
geometry.  The trigger response was also realistically simulated, and
systematic uncertainties were estimated. Next-to-leading order cross
sections were used whenever available, and used to normalise simulated
event samples.

\subsection{Higgs boson decay to four leptons:  $H \rightarrow Z Z^{(*)} \rightarrow 4 l$ ($4e$, $4\mu$ or $2e2\mu$)}

\begin{figure*}[t]
\centering
\includegraphics[width=62mm]{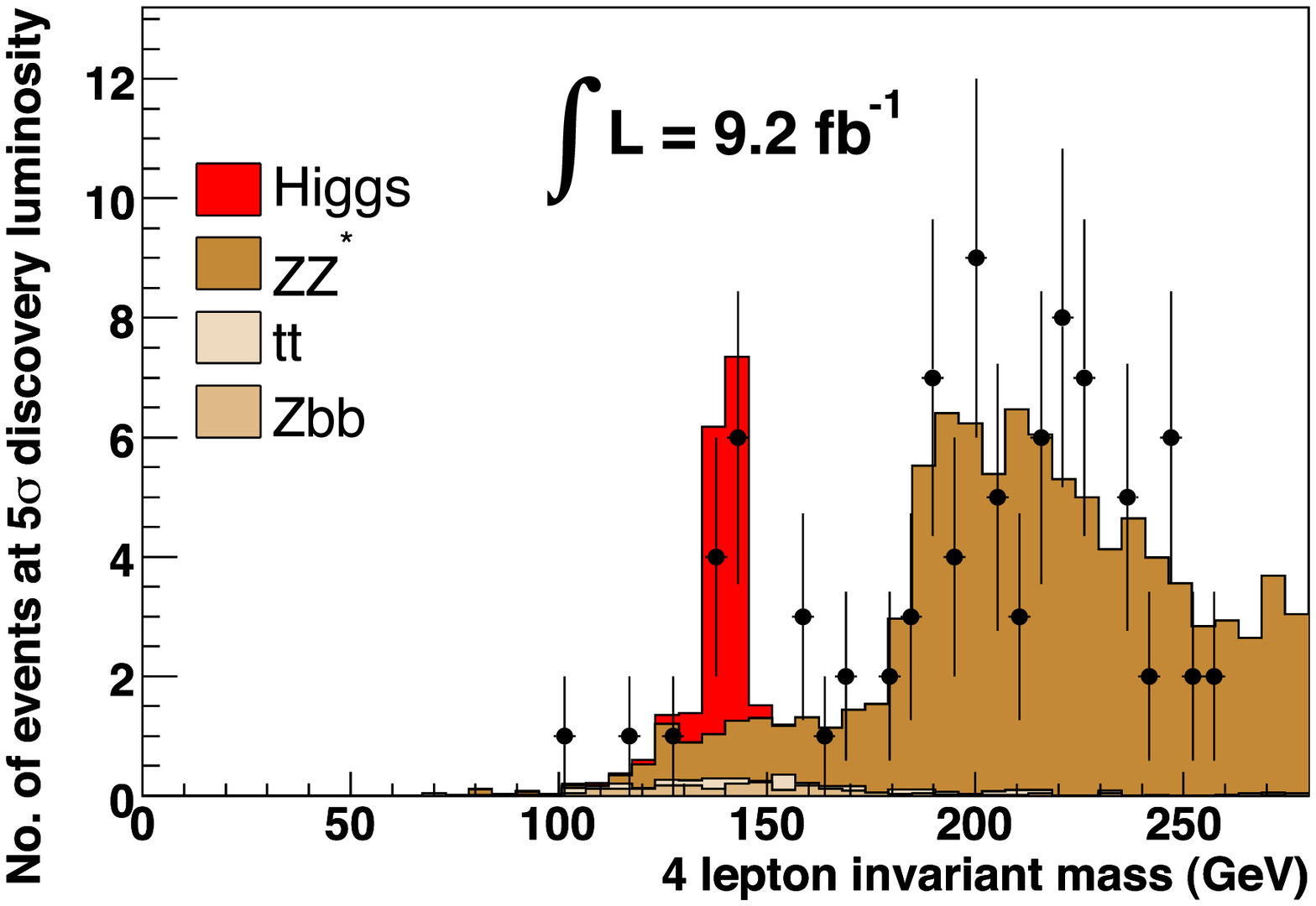}
\includegraphics[width=62mm]{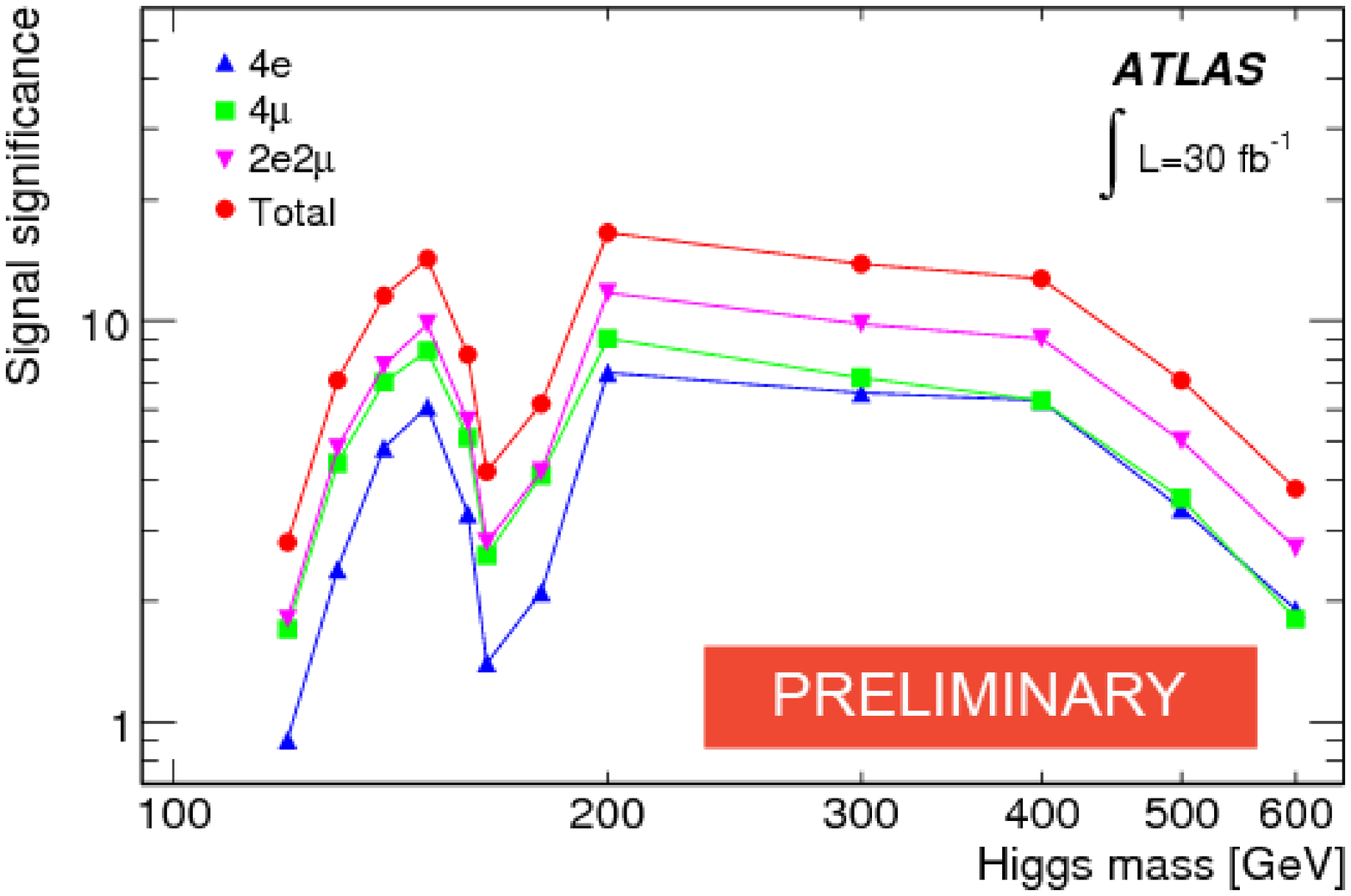}
\includegraphics[width=48mm]{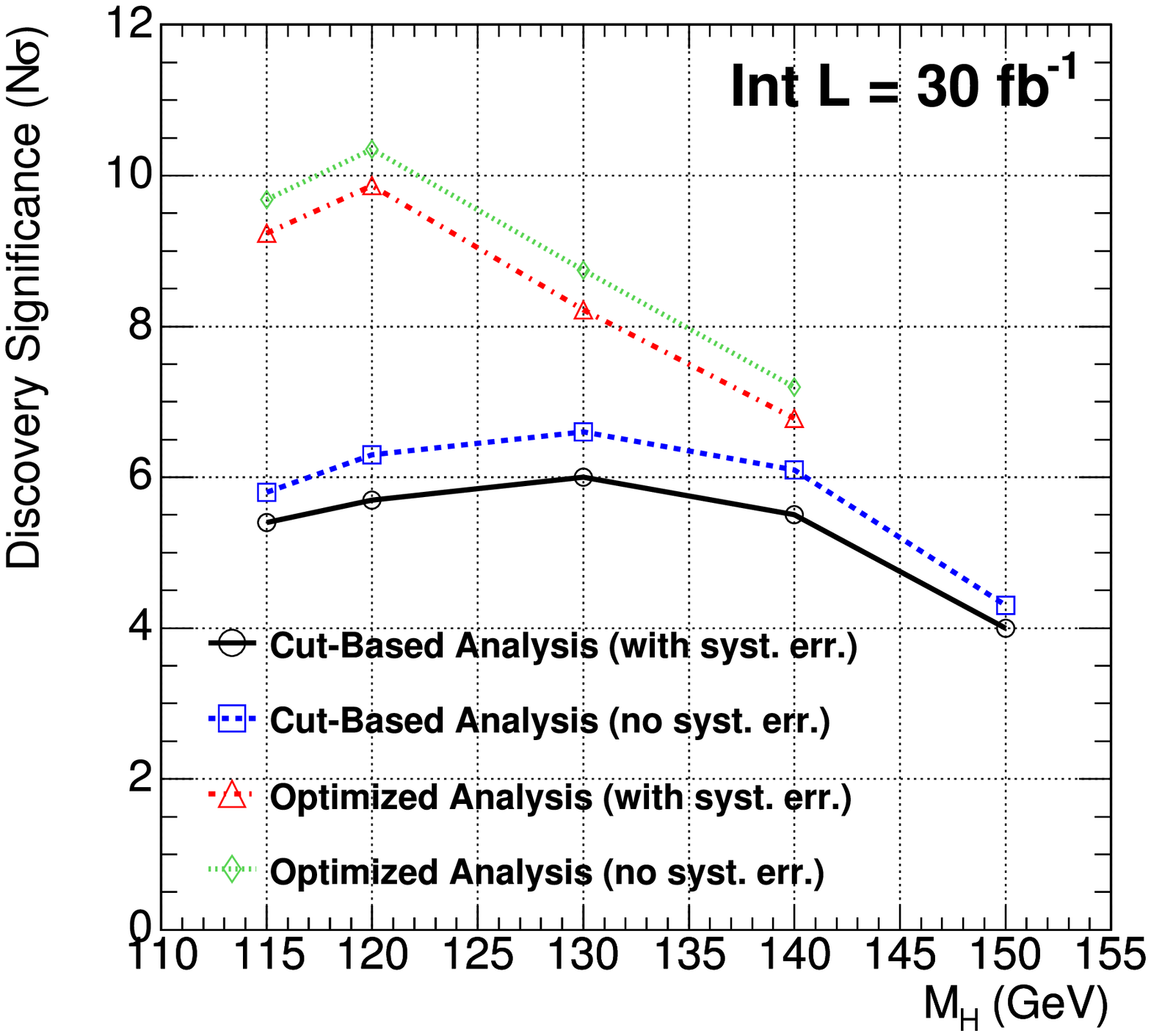}
\caption{Left: ATLAS sensitivity in the $H\rightarrow 4l$ channel for
  $30\,fb^{-1}$~\cite{ref:cscbook} (background estimation systematic
  uncertainties are included); centre: four-lepton invariant mass
  distribution in the $H \rightarrow Z Z^{(*)} \rightarrow 2e2\mu$
  channel in CMS~\cite{ref:cmstdr2}; right: CMS sensitivity in $H
  \rightarrow \gamma\gamma$ channel for
  $30\,fb^{-1}$~\cite{ref:cmstdr2}.} \label{fig:gg_4l}
\end{figure*}

This channel provides excellent sensitivity for a wide range of $m_{H}$
above $130 \,\rm{GeV}/c^2$, except for the interval between $2m_W$ and
$2 m_{Z}$, where the Higgs boson branching ratio is dominated by $H
\rightarrow W^+W^-$. The main background is $pp \rightarrow ZZ^{(*)}
\rightarrow 4l^{\pm}$. Other backgrounds, such as $Zb\bar{b}$, $ZW$,
$t\bar{t}$, and $Z+X$ are effectively suppressed by the analysis event
selection. Both CMS and ATLAS rely on selecting events which contain
pairs of electrons or muons with opposite charge. At least one $Z^0$
is expected to be on mass shell, and the two-lepton invariant mass is
used to reject fake and misidentified leptons.  A clear four-lepton
mass peak above a flat background is expected in this channel (see
figure~\ref{fig:gg_4l}, left), thus allowing a good background estimation
from the peak sidebands. Two factors are especially significant for
analyses of this channel: the lepton reconstruction efficiency and the
invariant-mass resolution.  Both experiments show similar experimental
sensitivities in this channel, with the total signal significance for
each experiment expected to be in excess of 10 for an integrated
luminosity of $30 \,fb^{-1}$ (see figure~\ref{fig:gg_4l}, centre).

\subsection{Higgs boson decay to two photons: $H \rightarrow \gamma \gamma$}

In spite of the low Higgs boson branching ratio to two photons (around
0.2\% for $m_{H}=120\,\rm{GeV}/c^{2}$), this channel shows very good
sensitivity in a range of $m_{H}$ between the LEP limit and $m_{H}
\sim 140 \, \rm{GeV}/c^{2}$.  The main irreducible background is the
direct $\gamma \gamma$ production. Signal appears as a sharp peak over
a smoothly falling di-photon invariant mass distribution.  Reducible
instrumental backgrounds such as $\gamma+jet$ and multijet production,
where one or more particles are misidentified as photons, are
important due to the large QCD cross sections.  Both experiments need
to take photon conversions into account, since $\sim 60\%$ of Higgs events
contain at least one converted photon.

Both collaborations classify the selected events into several
categories, according to the event kinematics or topology. In this
way, the event selection can be optimized, in order to maximize the
background discrimination without degrading the efficiency for
selecting signal events. In ATLAS, a significance of about $3.5\sigma$
is expected for $10\,fb^{-1}$, assuming
$m_{H}=120\,\rm{GeV}/c^{2}$. With an optimized analysis employing a
neural network, CMS expects up to $10\sigma$ after collecting
$30\,fb^{-1}$ and a $5\sigma$ after accumulating $8\,fb^{-1}$ (see
figure~\ref{fig:gg_4l}, right).  Analyses are also being developed to
target the associated production channels $ZH$, $WH$ and $t\bar{t}H$.

\subsection{Higgs boson decay to a tau pair: $H \rightarrow \tau^+ \tau^-$}

This channel currently provides one of the best sensitivities at low
$m_H$. Analyses rely on the topology of VBF events to provide
additional rejection against SM backgrounds. In this process, the
Higgs boson is radiated by $W^{\pm}$ or $Z^0$ bosons exchanged between
the interacting partons.  Due to the lack of colour flow between the
two interacting partons, the characteristic event topology consists of
two relatively forward jets with a rapidity gap in between, containing
little hadronic activity.  Both ATLAS and CMS investigate this channel
for all final states, in which at least one tau decays to an electron
or muon plus neutrinos.  The final state where both taus decay
hadronically was also investigated by ATLAS and proved to be feasible,
but there is as yet no detailed sensitivity estimates. The dominant
background in all cases is $Z+jets$ with the $Z$ boson decaying to
$\tau^+\tau^-$. Both collaborations have developed data-driven methods
to estimate these backgrounds. Other backgrounds are $W+N$ jets,
$t\bar{t}$, and di-jet events.

The $\tau^+\tau^-$ invariant mass reconstruction requires an
approximation, in which the $\tau^{\pm}$ is assumed to be collinear
with the visible lepton (in $\tau \rightarrow e$ or $\tau \rightarrow
\mu$ decays). The resulting Higgs mass resolution can be observed in
figure~\ref{fig:tau_ww} (left).  The expected sensitivity for this channel,
currently reaches up to $5\sigma$ for $30 \,fb^{-1}$.

\subsection{Higgs boson decay to a W boson pair: $H \rightarrow W^+W^-\rightarrow l\nu l\nu$ ($l = e^{\pm}$ or $\mu^{\pm}$)}

\begin{figure*}[t]
\centering
\includegraphics[width=47mm]{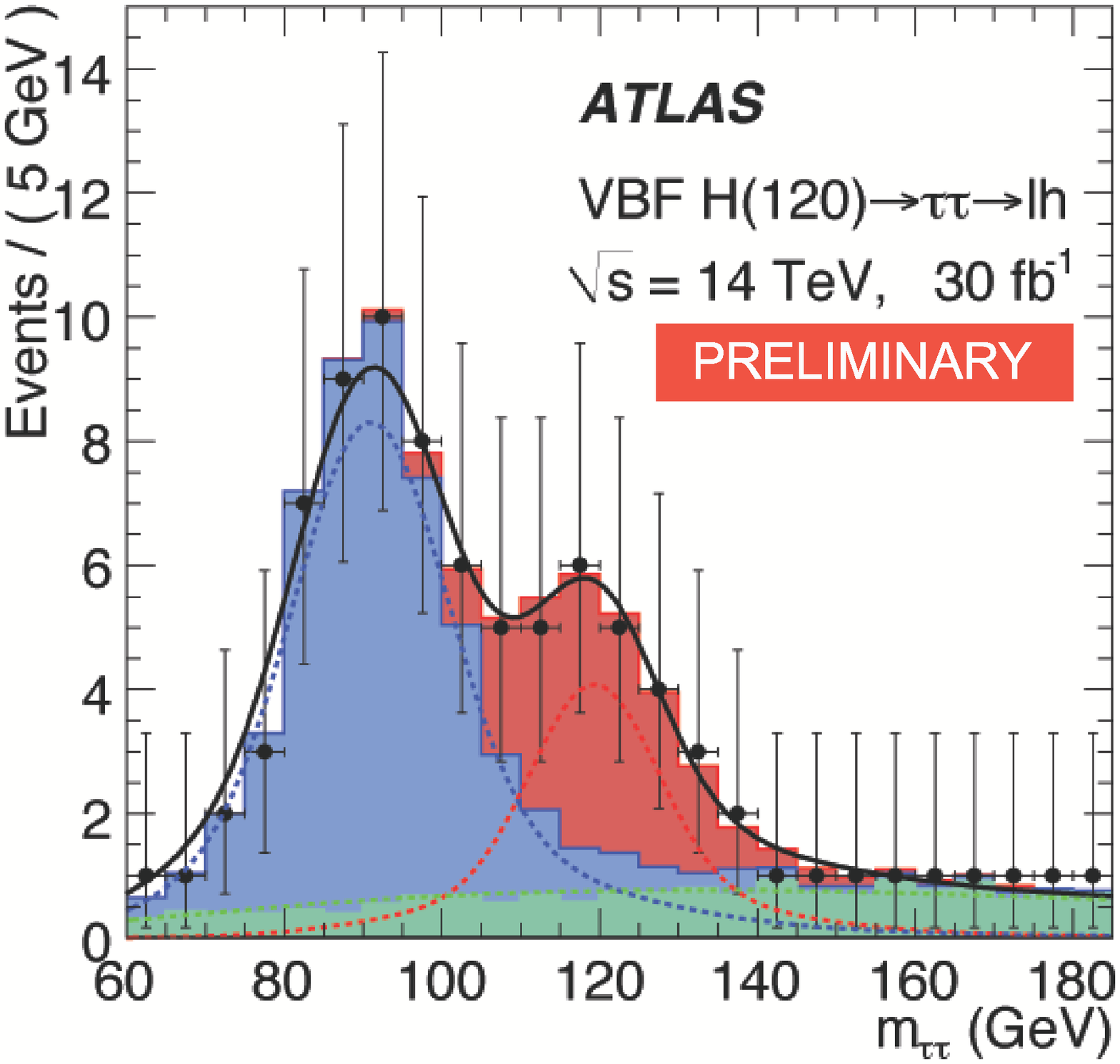}
\includegraphics[width=51mm]{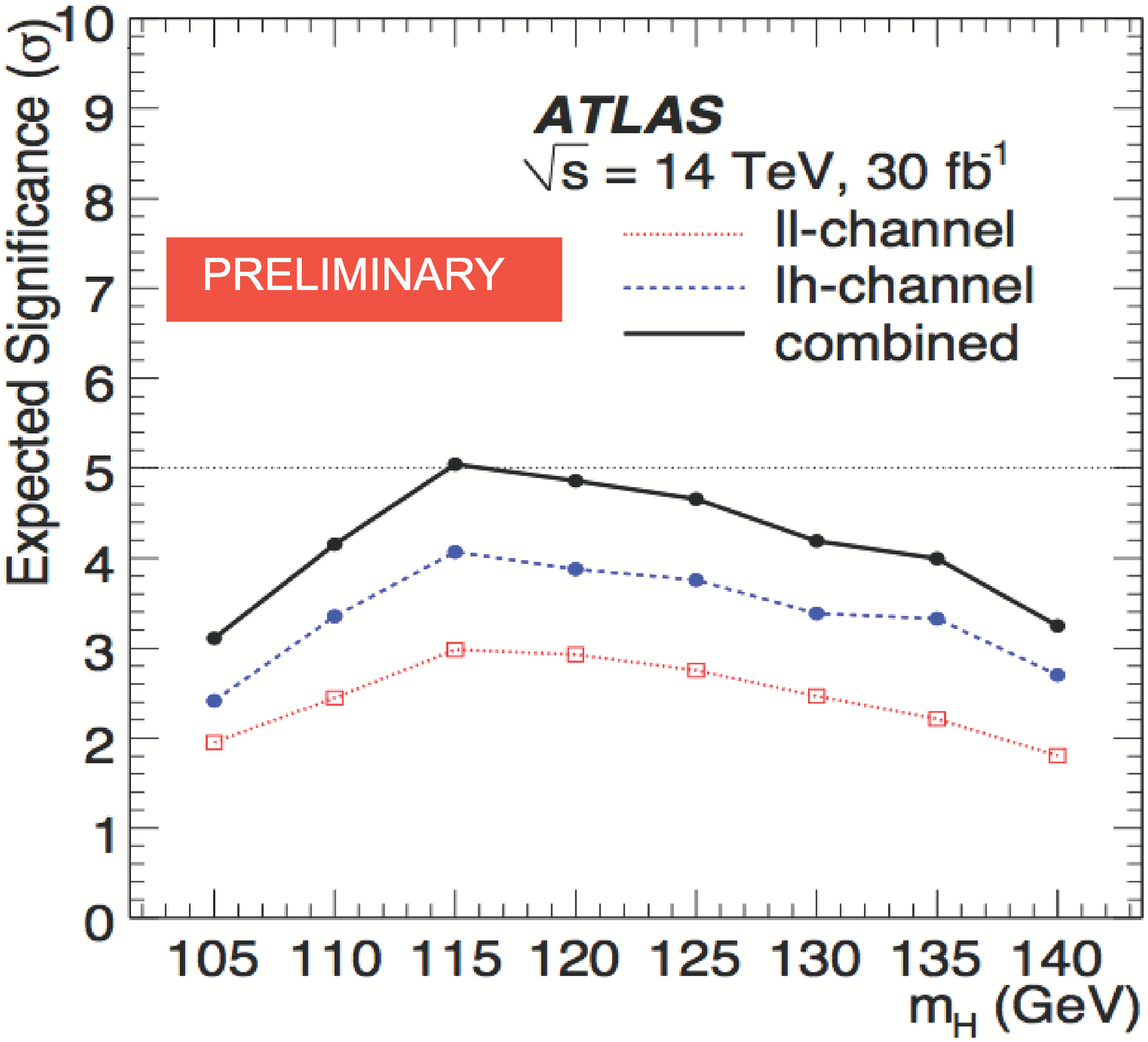}
\includegraphics[width=72mm]{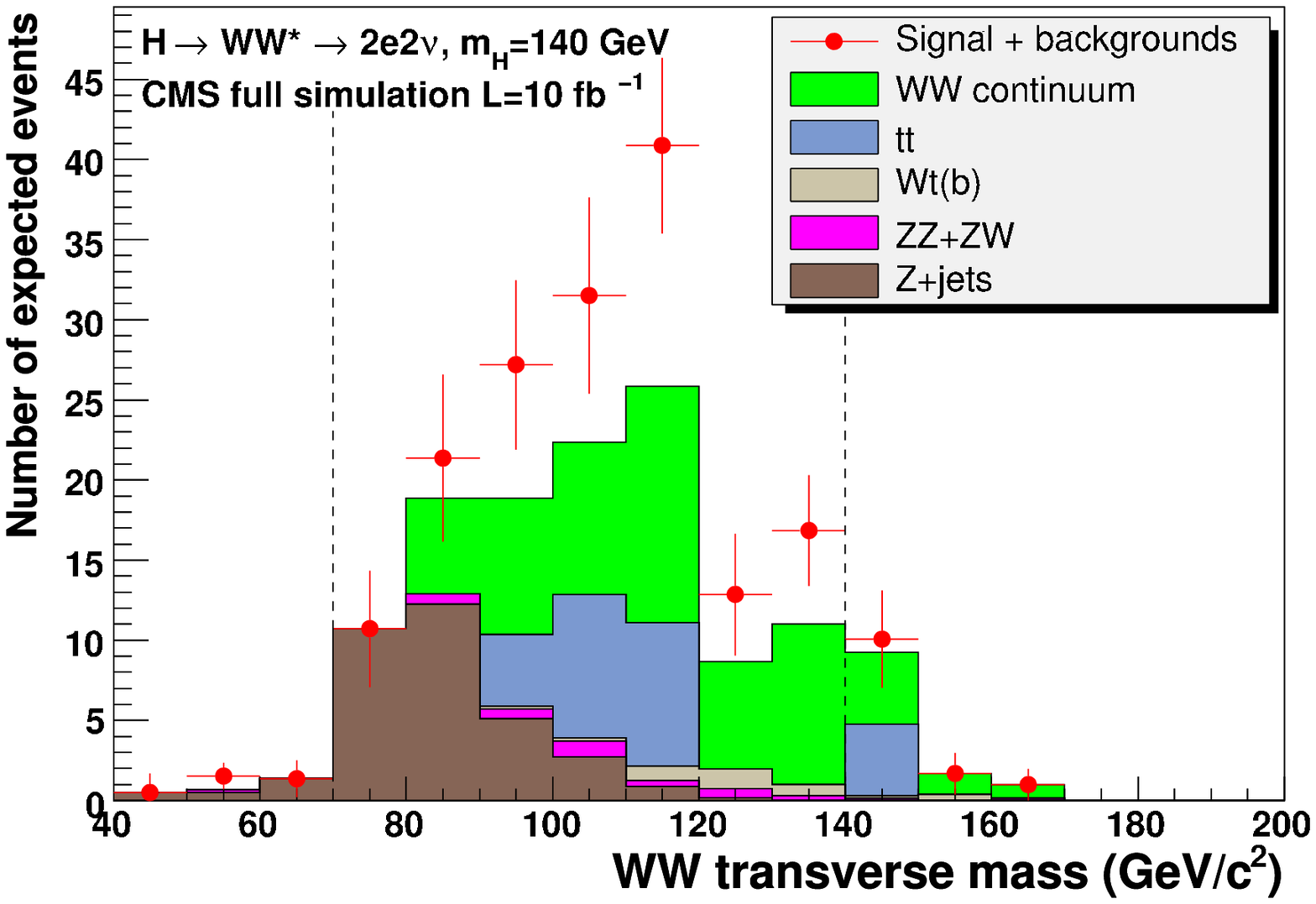}
\caption{Left: $\tau^+\tau^-$ invariant mass distribution in the $H
  \rightarrow \tau^+\tau^-$ channel in ATLAS with one $\tau$ decaying
  hadronically and $m_H=120 \,\rm{GeV}/c^2$~\cite{ref:cscbook};
  centre: expected sensitivity for the $H\rightarrow \tau^+\tau^-$
  channel in ATLAS (systematic uncertainties due to the background
  estimation are included)~\cite{ref:cmstdr2}; right: transverse mass
  in the $H \rightarrow W^+W^-\rightarrow l\nu l\nu$ in the CMS
  analysis~\cite{ref:cmstdr2}.} \label{fig:tau_ww}
\end{figure*}

The Higgs boson decay to a $W^+W^-$ pair provides the most sensitive
search channel in the mass range $2 m_W < m_H < 2 m_Z$, where this
decay mode has a branching ratio above 95\%. Analyses have
concentrated on final states containing electrons or muons from the
$W^{\pm}$ decay. Contrary to the remaining channels, though, the
presence of high transverse momentum neutrinos makes it unfeasible to
obtain a Higgs mass peak. The transverse mass~\cite{ref:cmstdr2} can
still be calculated and used in the event selection (see
figure~\ref{fig:tau_ww}, right). Analyses in this channel then need to rely on
a very good knowledge of the background shape and normalisation.

The dominant backgrounds come from events containing a W-boson pair,
most importantly $W^+W^-$ and $t\bar{t}$ production. The $H\rightarrow
W^+W^-$ decay creates a correlation between the W spins in the Higgs
reference frame which translates into an angular correlation between
the leptons emitted in the W-bosons decay. This is exploited to
suppress the $W^+W^-$ background. The $t\bar{t}$ background can be
effectively suppressed by a veto on central jets (jets with a small
pseudorapidity value). Separate analyses are performed for the cases
where there are no high transverse momentum jets present in the event
and where there are two additional jets, directed at the gluon fusion
and the VBF production modes, respectively. For a Higgs boson mass
close $160 \,\rm{GeV}/c^2$ a sensitivity in excess of $10\sigma$ is
expected for each experiment for $30 \,fb^{-1}$ of integrated
luminosity. Additional channels being explored are the VBF $H
\rightarrow W^+W^- \rightarrow l\nu +2jets$ and the associated
production channels $W^{\pm} H$ and $t\bar{t}H$.

\section{Summary and Outlook}

\begin{figure*}[t]
\centering
\includegraphics[width=67mm]{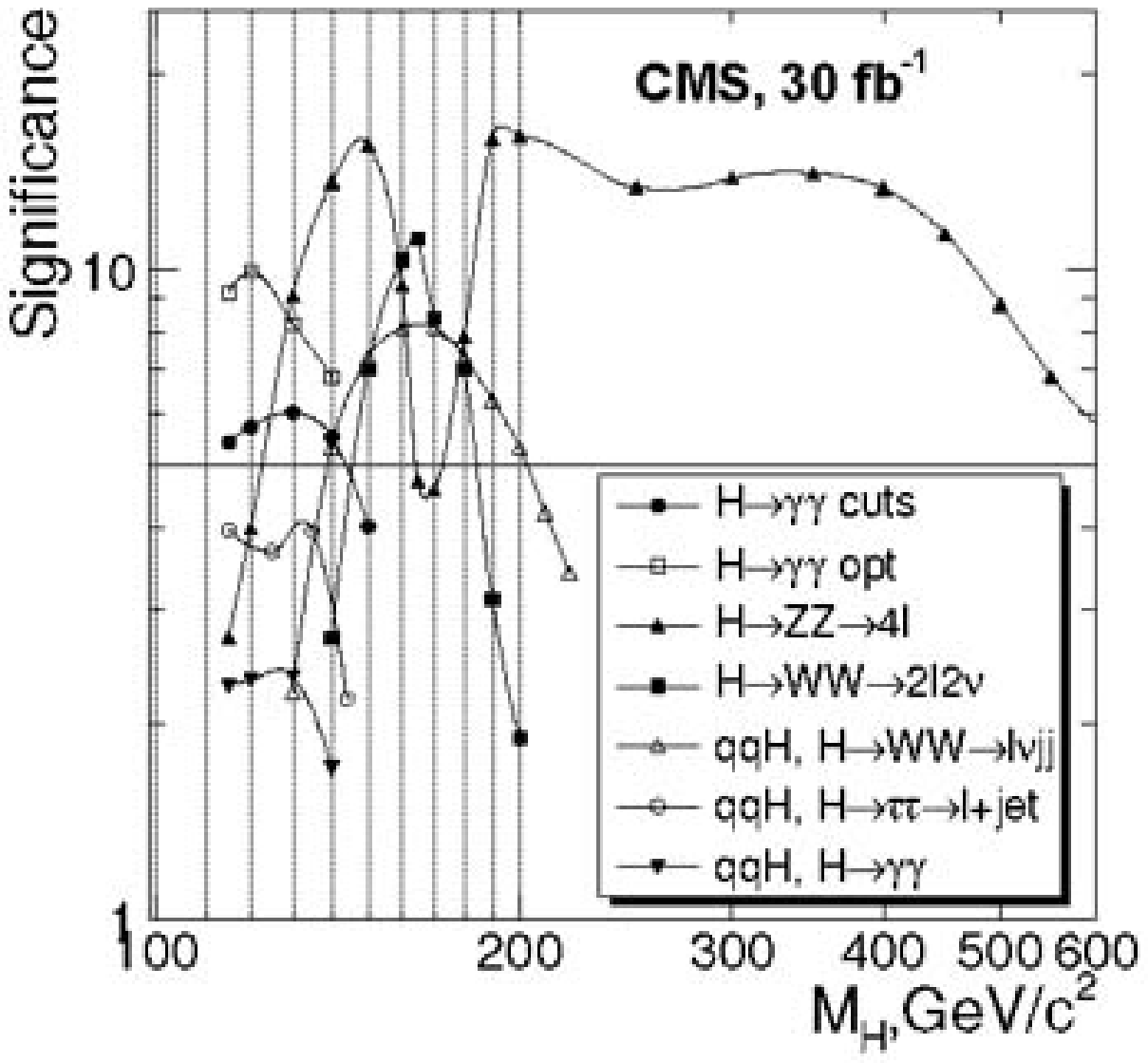}
\includegraphics[width=70mm]{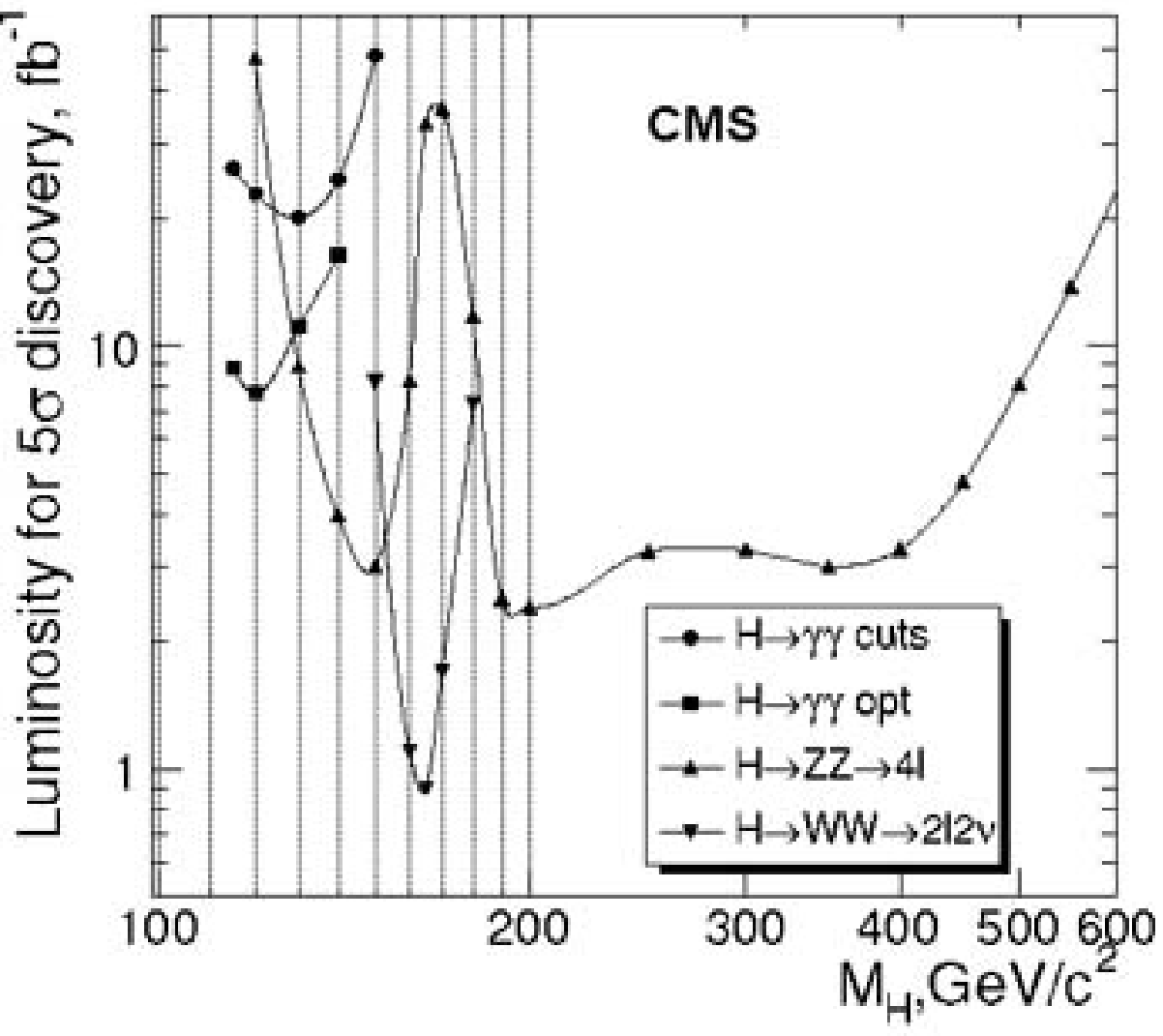}
\caption{SM Higgs discovery potential for CMS assuming an integrated
  luminosty of $30\,fb^{-1}$ (left) and integrated luminosity needed
  for a $5\sigma$ Higgs discovery, versus Higgs boson
  mass~\cite{ref:cmstdr2}).} \label{fig:sensitivity}
\end{figure*}

The expected sensitivity of ATLAS and CMS to a SM Higgs boson was
evaluated by both collaborations. Analyses in both experiments show
similar sensitivity in most channels.  The sensitivity for SM Higgs
discovery in the various channels is illustrated by
figure~\ref{fig:sensitivity} for the CMS case. An integrated
luminosity of less than $10 \,fb^{-1}$ should be enough to make a
$5\sigma$ discovery in the mass range above the LEP limit. Preliminary
ATLAS studies indicate that an integrated luminosity of $2\,fb^{-1}$
would be enough to exclude the Higgs boson in the range $121
\rm{GeV}/c^2 \lesssim m_H \lesssim 460 \rm{GeV}/c^2$ at 95\%
confidence level. Other analyses, not described here, may contribute
to further enhancements of the sensitivity.

\begin{acknowledgments}
  The author would like to express his thanks to ATLAS and CMS
  colleagues. Special thanks to Chiara Mariotti, Yves Sirois, Louis Fayard,
  Aleandro Nisati, Ketevi Assamagan, Eilam Gross, and Glen Cowan, and
  to many other colleagues who have kindly contributed with work and
  material for this paper.
\end{acknowledgments}

\end{document}